\documentclass[12pt,preprint]{aastex}
\usepackage{lscape}

\usepackage{graphicx}
\graphicspath{{./figure/}}

\newcommand{\Fermi}{{\textit{Fermi}}}
\newcommand{\fermi}{{\textit{Fermi}}}
\newcommand{\GRB}{{GRB~090926A }}
\newcommand{\Tz}[1]{{$T_0+#1$\,s}}

\slugcomment{Accepted by ApJ, January $9^{th}$ 2011}

\begin{document}

%\title{Fermi Observations of \GRB}
\title{Detection of a spectral break in the extra hard component of GRB~090926A}

%\author{Contact Authors:\\
%Johan Bregeon (johan.bregeon@pi.infn.it),
%Adam Goldstein (adam.m.goldstein@nasa.gov),
%Rob Preece (Rob.Preece@nasa.gov),
%Hiromitsu Takahashi (hirotaka@hep01.hepl.hiroshima-u.ac.jp),
%Kenji Toma (toma@astro.psu.edu),
%Takeshi Uehara (uehara@hep01.hepl.hiroshima-u.ac.jp)}

% add this below to have "all" authors
%\title{Detection of a spectral cutoff in the extra hard component from GRB 090926A\\Author list created Thursday 24 Jun 2010 07:23 PDT}
\author{
M.~Ackermann\altaffilmark{2}, 
M.~Ajello\altaffilmark{2}, 
K.~Asano\altaffilmark{3}, 
M.~Axelsson\altaffilmark{4,5,6}, 
L.~Baldini\altaffilmark{7}, 
J.~Ballet\altaffilmark{8}, 
G.~Barbiellini\altaffilmark{9,10}, 
M.~G.~Baring\altaffilmark{11}, 
D.~Bastieri\altaffilmark{12,13}, 
K.~Bechtol\altaffilmark{2}
% ...
}
% ...
\altaffiltext{2}{W. W. Hansen Experimental Physics Laboratory, Kavli Institute for Particle Astrophysics and Cosmology, Department of Physics and SLAC National Accelerator Laboratory, Stanford University, Stanford, CA 94305, USA}
\altaffiltext{3}{Interactive Research Center of Science, Tokyo Institute of Technology, Meguro City, Tokyo 152-8551, Japan}
\altaffiltext{4}{Department of Astronomy, Stockholm University, SE-106 91 Stockholm, Sweden}
\altaffiltext{5}{Lund Observatory, SE-221 00 Lund, Sweden}
\altaffiltext{6}{The Oskar Klein Centre for Cosmoparticle Physics, AlbaNova, SE-106 91 Stockholm, Sweden}
\altaffiltext{7}{Istituto Nazionale di Fisica Nucleare, Sezione di Pisa, I-56127 Pisa, Italy}
\altaffiltext{8}{Laboratoire AIM, CEA-IRFU/CNRS/Universit\'e Paris Diderot, Service d'Astrophysique, CEA Saclay, 91191 Gif sur Yvette, France}
\altaffiltext{9}{Istituto Nazionale di Fisica Nucleare, Sezione di Trieste, I-34127 Trieste, Italy}
\altaffiltext{10}{Dipartimento di Fisica, Universit\`a di Trieste, I-34127 Trieste, Italy}
\altaffiltext{11}{Rice University, Department of Physics and Astronomy, MS-108, P. O. Box 1892, Houston, TX 77251, USA}
\altaffiltext{12}{Istituto Nazionale di Fisica Nucleare, Sezione di Padova, I-35131 Padova, Italy}
\altaffiltext{13}{Dipartimento di Fisica ``G. Galilei", Universit\`a di Padova, I-35131 Padova, Italy}
% ...

\begin{abstract}
We report on the observation of the bright, long gamma-ray burst,
GRB~090926A, by the Gamma-ray Burst Monitor (GBM) and Large Area
Telescope (LAT) instruments on board the \Fermi\ Gamma-ray Space Telescope. 
GRB~090926A shares several features with other bright LAT bursts. In
particular, it clearly shows a short spike in the light curve that is
present in all detectors that see the burst, and this in turn suggests
that there is a common region of emission across the entire \Fermi\
energy range.
In addition, while a separate high-energy power-law component has already been observed in other GRBs, here we report for the first time the detection with good significance of a high-energy spectral break (or cutoff) in this power-law component around 1.4~GeV
in the time-integrated spectrum.
If the spectral break is caused by opacity to electron-positron pair production within the source, then this observation allows us to compute the bulk Lorentz factor for the 
outflow, rather than a lower limit.
\end{abstract}

\keywords{gamma rays: bursts}

%\documentclass[12pt,preprint]{aastex}
%\documentclass{emulateapj}
%\usepackage{apjfonts}
%\usepackage{amsmath}
%
%

%\begin{document}

\section{Introduction}\label{intro}

{Gamma-Ray Bursts (GRBs)} are the most energetic transients in the universe. 
The first brief and intense flash, the so-called prompt emission,
has been observed in the X-ray and gamma-ray bands, while subsequent
long-lived afterglow emission has so far been observed mainly at energies in the X-ray band and below.
The prompt emission is thought to be produced in an ultra-relativistic outflow,
but its detailed emission mechanism has been a long-standing problem.
It has been widely believed that the afterglow is the synchrotron emission from the forward shock
that propagates in the external medium, but {\it Swift} observations 
have pointed out some difficulties in this model 
\citep[for recent reviews,][]{2007ChJAA...7....1Z, 2006RPPh...69.2259M}.
The study of the gamma-ray emission in the GeV energy range is expected to give
us important information on these issues and even on the nature of the progenitors
and the ultra-relativistic outflows of GRBs \citep{2009ApJ...701.1673B,2008arXiv0810.0520F,2008FrPhC...3..306F}.
%High-energy emission from GRBs was first detected by the Energetic Gamma-Ray Experiment Telescope (EGRET),covering the energy range from 30 MeV to 30 GeV, on board the Compton Gamma-Ray Observatory (CGRO; 1991-2000), but EGRET detected only 5 GRBs \citep{1995Ap&SS.231..187D,Hurley:94,Gonzalez:03}. 
%The Italian experiment Astro-rivelatore Gamma a Immagini LEggero (AGILE; 2007-) has 
%detected a number of high-energy photons with energies up to 300 MeV from 
%GRB~080514B and GRB~090510 \citep{2008A&A...491L..25G, 2009arXiv0908.1908G}.
%The origin of the high-energy photons in these GRBs is not clear, and 
%more data are needed in order to put some conclusive constraints on the mechanism
%of the GRB high-energy emission.

The {\it Fermi} Gamma-ray Space Telescope hosts two instruments, the Large Area Telescope
\citep[LAT, $20\;{\rm MeV}$ to more than $300\;$GeV;][]{2009ApJ...697.1071A} and 
the Gamma-ray Burst Monitor \citep[GBM, $8\;{\rm keV}$--$40\;{\rm MeV}$;][]{Meegan_GBM}, which together are 
capable of measuring the spectral parameters of GRBs across seven decades in energy.
Since the start of science operations in early August 2008, 
the {\it Fermi} LAT has significantly detected 16 GRBs.
These events, including the very bright long-duration and short-duration bursts 
GRB~080825C, GRB~080916C, GRB~081024B, GRB~090510, and GRB~090902B, 
have revealed many important, seemingly common, features of GRB GeV emission
\citep{Fermi...GRB090902B, 2009Sci...323.1688A, Nature..GRB090510, Fermi..GRB080825C, Fermi..GRB081024B, 2010ApJGRB090510, 2010ApJGRB090217}:
(1) {the} GeV emission onsets of many LAT GRBs are delayed with respect to the MeV emission onsets;
(2) {some LAT GRBs} have extra hard components apart from the canonical Band function \citep{Band:93}, which typically peaks in $\nu F_{\nu}$ between around 100 keV--1 MeV;
(3) the GeV emission lasts longer than the prompt MeV emission, showing power-law temporal decays at late times.

%In addition to these properties, the LAT has not detected any spectral break or cutoff, from any of these previous bursts.
%The prompt high-energy photons may be absorbed via creating electron-positron
%pairs with low-energy radiation in the emitting region $(\gamma\gamma \to e^+ e^-)$.
%Since the opacity depends on the unknown bulk Lorentz factor and the size of
%the emitting region, the non-detections of the internal absorption signatures have
%put constraints on these physical parameters for some GRBs, which are crucial for
%understanding the emission mechanism
%\citep{Fermi...GRB090902B,2009Sci...323.1688A,Nature..GRB090510,2010ApJGRB090510}.

In this paper, we report on the analysis of the bright, long
GRB~090926A detected by \Fermi\ LAT/GBM.
The light curve of this burst above $100\;$MeV shows a sharp spike with a width of $0.15\;$s, fast variability that we use to constrain the origin of the high-energy photons within the spike. Furthermore, from the detection of a break in the $>$\,100\,MeV gamma-ray spectrum, we derive constraints on the bulk Lorentz factor and the distance of the emitting region from the central source. Section \ref{section:obs} summarizes the detections of GRB~090926A by the GBM and the LAT, and the follow-up observations.
Section \ref{section:lightcurves} presents the light curves of the prompt emission as seen by both instruments and describes a sharp pulse seen in all detectors.
In section \ref{section:ana}, we detail the spectral analysis of the burst through time-resolved spectroscopy,  the measurement of a break in the extra-component, and the extended emission found in the LAT data out to $4.8\;$ks after the trigger.
These last two points are at the center of the physical interpretation of the observations that is developed in section \ref{sec:discussion}. Throughout this paper, we adopt a Hubble constant of $H_0 = 72\;{\rm km}\;{s}^{-1}\;{\rm Mpc}^{-1}$ and cosmological parameters of $\Omega_{\Lambda} = 0.73$ and $\Omega_{M} = 0.27$.

%Finally we summarize our results in section \ref{sec:conclusions}.

%\begin{thebibliography}{99}
%\end{thebibliography}

%\end{document}

\section{Observations}\label{section:obs}
At 04:20:26.99 (UT) on Sept. 26, 2009 
%(MET, mission elapsed time is $T_0=$ 
%275631628.98), 
(hereafter $T_0 = 275631628.98$\,s mission elapsed time),
the \fermi~Gamma-ray Burst Monitor (GBM) triggered on 
and localized the long \GRB at (RA, Dec) = (354.5$^\circ$, $-64.2^\circ$), 
in J2000 coordinates \citep{2009GCN..9933....1B,LAT_090926A}.
This position was $\sim$52$^\circ$ with respect to the LAT boresight
at the time of the trigger and well within the 
field of view.
An Autonomous Repoint Request (ARR) was generated, but the spacecraft
initially remained in survey mode as the Earth avoidance
angle condition was not satisfied by the burst pointing direction. The on-board GBM position of \GRB was occulted by the Earth at roughly \Tz{500} until it rose above the horizon at approximately \Tz{3000}.
At that time, the spacecraft slewed to \GRB and kept it close to the center of the LAT field of view until \Tz{18000}, though the source location was occulted by the Earth several times over that time period.

Emission from \GRB was evident in the \fermi~LAT raw trigger event rates, and
the number of LAT events ($\sim200$ photon candidates above 100 MeV) is comparable to that of the other bright LAT bursts, GRB~080916C, GRB~090510 and GRB~090902B.
The increase in the photon count rate during the prompt phase is spatially and
temporally correlated with the GBM emission with high significance, and extended emission is observed until \Tz{4800}. 
The best LAT on-ground localization is (RA, Dec) = (353.56$^\circ$, $-66.34^\circ$), with a 90\% containment radius of 0.07$^\circ$ (statistical; 68\%
containment radius: 0.04$^\circ$, systematic
error is less than 0.1$^\circ$) and is consistent with the XRT localization.

Indeed, based upon the GCN report issued for the LAT detection, 
a {\it Swift} TOO observation was performed, and an afterglow for \GRB
was detected with
XRT and UVOT at $T_0+47\,$ks and localized at (RA, Dec) = (353.40070$^\circ$,
-66.32390$^\circ$) with an uncertainty of 1.5 \arcsec 
(90\% confidence) \citep{2009GCN..9936....1G}.
VLT observations determined a redshift for \GRB of $z=2.1062$, using 
the X-shooter spectrograph \citep{2009GCN..9942....1M}. 
Suzaku/WAM and Skynet/PROMPT also detected the soft gamma-ray prompt and 
optical afterglow emission, respectively \citep{2009GCN..9951....1N}.

\section{Light Curves}
\label{section:lightcurves}
\begin{figure}
  \centering
  \includegraphics[scale=0.6]{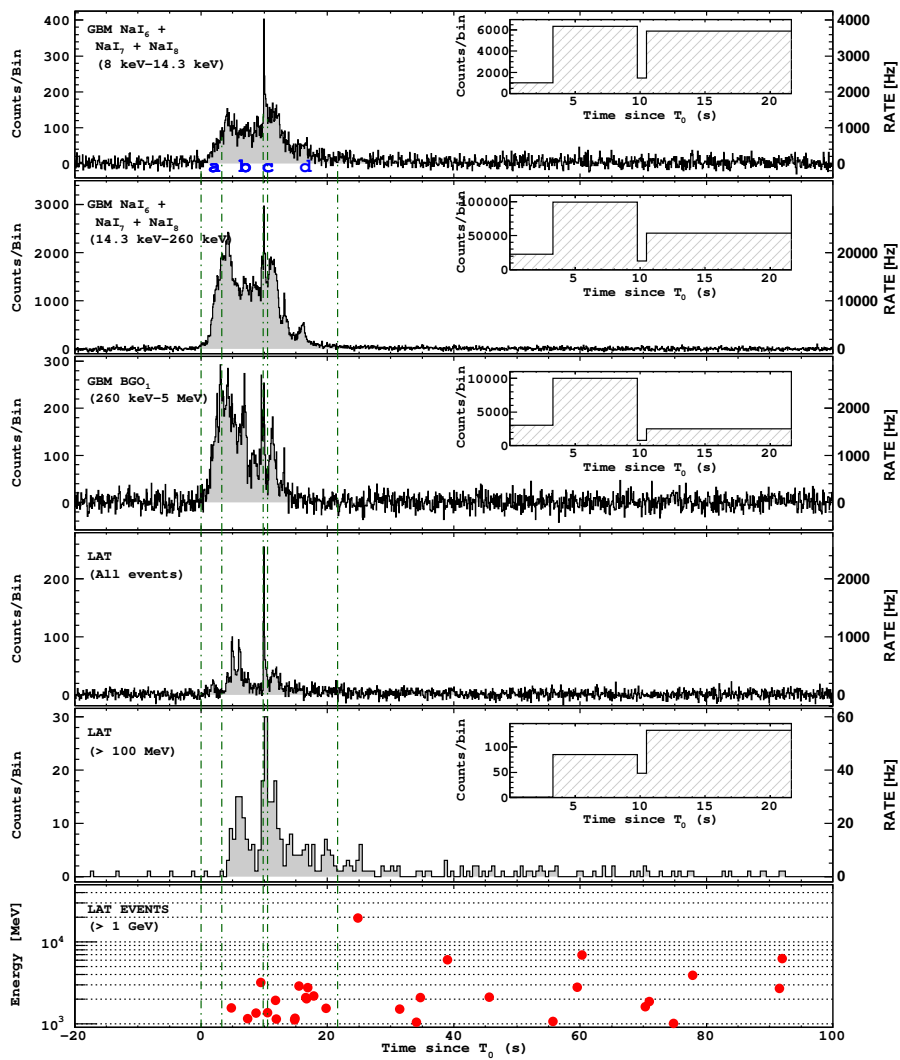}
  \caption{GBM and LAT light curves for the gamma-ray emission of GRB~090926A. 
The data from the GBM NaI detectors were divided into soft (8--14.3\,keV) 
and hard (14.3--260\,keV) bands to reveal similarities between the light curve 
at the lowest energies and that of the LAT data.
The fourth panel shows all LAT events that pass the on-board GAMMA filter~\citep{2009ApJ...697.1071A}. 
The first four light curves are background-subtracted and are shown for 0.1\,s time bins.
The fifth and sixth panels show LAT data `transient' class events  
for energies $>$\,100\,MeV and $>$\,1\,GeV respectively, both using 0.5\,s time bins.
The vertical lines indicate the boundaries of the intervals 
used for the time-resolved spectral analysis, $T_0 + (0, 3.3, 9.8, 10.5, 21.6)$ $\,$s. 
The insets show the counts for each data set, 
binned using these intervals, to illustrate the numbers of counts considered 
in each spectral fit.}
  \label{Fig:prompt light curves}
\end{figure}

% Main light curve and time bins
In figure~\ref{Fig:prompt light curves}, we show the GBM and LAT light
curves in several energy bands.
The highest energy photon is a 19.6 GeV event, observed at \Tz{25} 
within 0.03$^\circ$ from the LAT position of GRB~090926A, well within the 68\% containment of the point spread function at that energy.
The light curves show that the onset of the LAT emission is delayed by 
3.3 s with respect to the GBM emission, similar to other LAT GRBs 
\citep{Fermi...GRB090902B, 2009Sci...323.1688A, Nature..GRB090510, Fermi..GRB080825C, Fermi..GRB081024B, 2010ApJGRB090510, 2010ApJGRB090217}. 
Detailed analysis of the GBM data results in a formal T90 
duration\footnote[1]{The T90 duration is the time over which the central 90\% of the counts between 50 and 300 keV have been accumulated.} 
\citep{Kouv93} of $13.1\pm0.2$ s, with  a start time of \Tz{2.2} and a stop time of \Tz{15.3}. The emission measured in the LAT above 100\,MeV has a similar duration; however, owing to the efficient background rejection applied to the LAT data, the signal is clearly visible in the light curve well after this time range.

\begin{figure}
  \centering
  \includegraphics[scale=0.6]{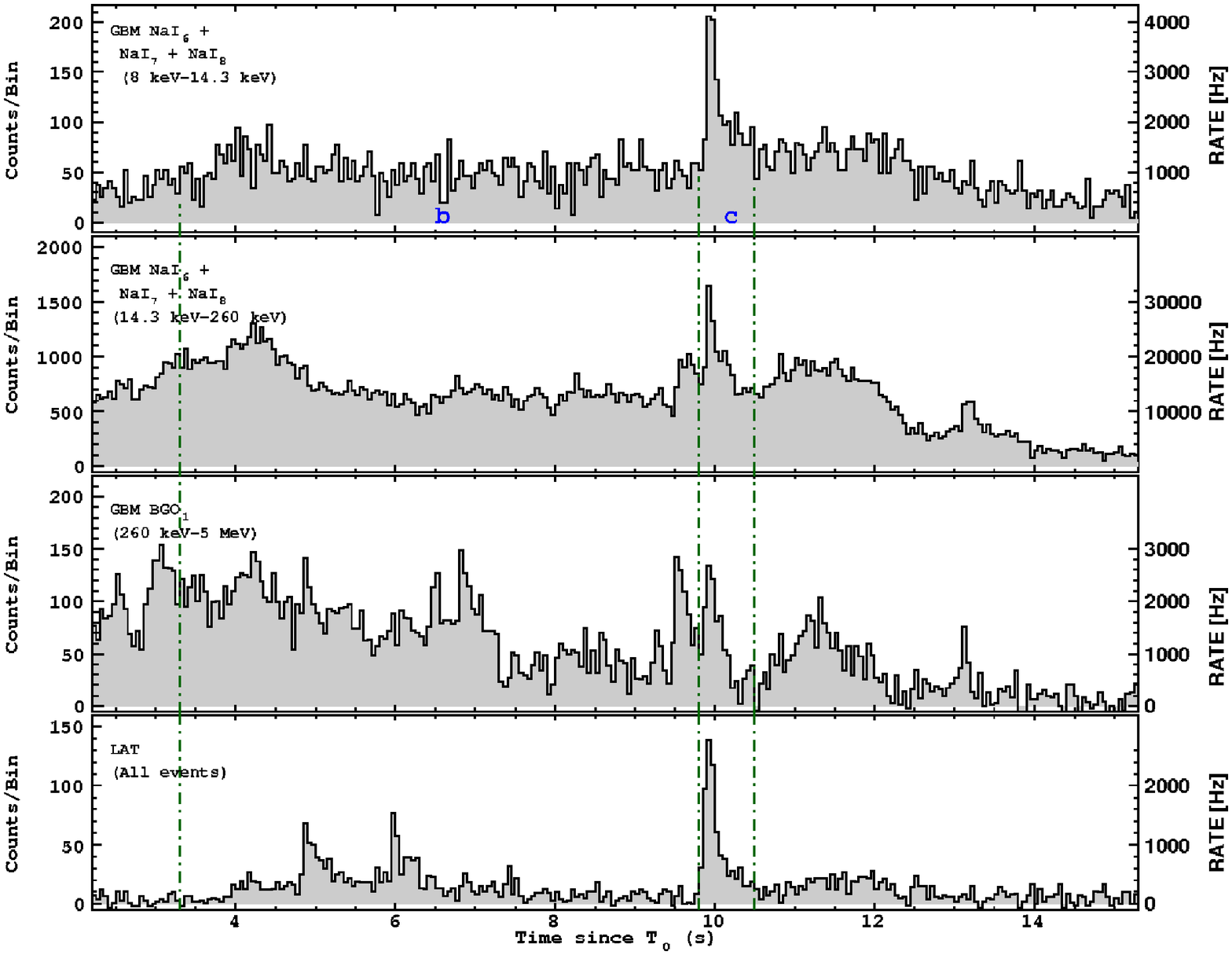}
  \caption{GBM and LAT light curves for the gamma-ray emission of \GRB with $0.05$\,s time binning for the core of the prompt phase. The vertical dashed lines at \Tz{9.8} and \Tz{10.5} define interval \emph{c} used in the spectral analysis,$\;$ 
  see section~\ref{section:10speak}.}
  \label{Fig:prompt light curves zoom}
\end{figure}

% Zoom around the 10 s peak
The time intervals chosen for spectroscopy are indicated by the vertical lines in 
figure~\ref{Fig:prompt light curves}, with boundaries at $T_0 + (0, 3.3, 9.8, 10.5, 21.6)$ s. The end of the last time interval at \Tz{21.6} was chosen somewhat arbitrarily as the end of the prompt phase, but we carefully verified that our results are not affected by a slightly different choice.
Figure~\ref{Fig:prompt light curves zoom} shows a zoom of some of the light curves between \Tz{2.2} and \Tz{15.3}, with a binning of 0.05\,s, and highlights the presence of the sharp peak seen in each of the NaI, BGO, and LAT light curves at \Tz{10}.
As seen on figure~\ref{Fig:prompt light curves zoom}, the peak is clearly in coincidence in all of the light curves, indicating a strong correlation of the emission
from a few keV to energies $>100$ MeV.  Because this peak is the only one evident at all energies, we chose to run a dedicated spectral analysis between \Tz{9.8} and \Tz{10.5} as described in section~\ref{section:10speak}.

%Variability 
We estimated the variability time scale using the full width at half maximum (FWHM) of the bright pulse seen around \Tz{10}.
A combination of exponential functions is used to fit 
the light curve as performed in \citet{Norris:96}.
The light curve fitting is performed for all bright NaI detectors (N6, N7, N8) 
with a 2$\,$ms time resolution. Two exponential functions are used to represent 
the weak and main bright peaks and include a quadratic function to fit the longer timescale variations. As a result, we obtain a FWHM of the main peak of 0.15 $\pm$ $0.01\,$s for the bright pulse.

% define extended emission time interval and refer to the correct section
%On time scales longer than the prompt phase, we define the extended emission as the time 
%period $T>$\Tz{21.6}.
%The extended emission analysis reveals a detection in the LAT for up to \Tz{300} with 
%very high significance, and a clear detection of further extended emission 
%to \Tz{4800}; see the details discussed in section~\ref{section:extended}.

\section{Spectral Analysis}\label{section:ana}

\subsection{LAT and GBM spectral fitting}
\label{section:rmfit}
\label{section:correlation}
\label{section:10speak}
We performed a time-integrated joint spectral analysis of the LAT and GBM data for the prompt phase defined as \Tz{3.3} to \Tz{21.6} in figure~\ref{Fig:prompt light curves}.
For the GBM, we used `Time Tagged Events' (TTE) data from the NaI detectors 6, 7, 8
and BGO detector 1. As in \cite{Fermi..GRB080825C}, background rates and errors are estimated during the prompt phase by fitting background regions of the light curve before and after the burst.  We derived our background estimates using the time intervals $[T_0-44; T_0-8]\,$s and $[T_0+36; T_0+100]\,$s
for the NaI detectors, and $[T_0-43; T_0-16]\,$s and $[T_0+43; T_0+300]\,$s for the BGO detector.
For the LAT, we extracted `transient' class data from an energy-dependent acceptance cone around the burst position, as described in \cite{Fermi..GRB080825C}, and considered front- and back-converting events separately \citep{2009ApJ...697.1071A}.  The data files for the analysis were prepared using the LAT \texttt{ScienceTools-v9r15p2} package, which is available from the \fermi\ Science Support Center (FSSC), and the \texttt{P6\_V3\_TRANIENT} response functions.\footnote{\texttt{http://fermi.gsfc.nasa.gov/ssc/}}
A synthetic background was derived for the LAT data using an empirical model of the rates expected for the position of the source in the sky 
and for the position and orientation of the spacecraft during the burst interval.

The joint spectral fitting of GBM and LAT data was performed using {\it rmfit} version 3.2 \citep{Kaneko:06,Fermi..GRB080825C}, which estimates the goodness-of-fit in terms of the Castor Statistic (C-STAT) to handle correctly the small number of events at the highest energies. The Castor statistic~\citep{Dorman2003} is similar to the Cash statistic~\citep{Cash1979} except for an offset that is constant for a particular dataset. A global effective area correction has been applied to the BGO data to match the model normalizations given by the NaI data; this correction is consistent with the relative uncertainties in the GBM detector responses. Uncorrected, this will normally cause a mismatch between the fitted model rates between the two types of detectors where they overlap in energy. Once the correction has been determined, it is held fixed throughout the calculation, since it reflects an uncertainty in the response rather than in the data. In this analysis, the NaI to BGO normalization factor was found to be 0.79.
For further details on the data extraction and spectral analysis procedures see our previous publications~\cite{Fermi..GRB080825C} and~\cite{Fermi..GRB081024B}.

Initially, we fitted a canonical Band function~\citep{Band:93} to the data and then found that adding an extra power-law component improved both the fit statistics and residuals. Table~\ref{tab:spec} summarizes the best-fit parameters and shows that the improvement in C-STAT for the (Band+PL) fit over the Band fit alone is 107.3, indicating a firm detection of the additional power-law component.  The parameters of the Band function are stable, and the power-law photon index of the additional component is $\lambda=-1.79\pm0.02$.

\begin{figure}
  \centering
  \includegraphics[scale=0.5]{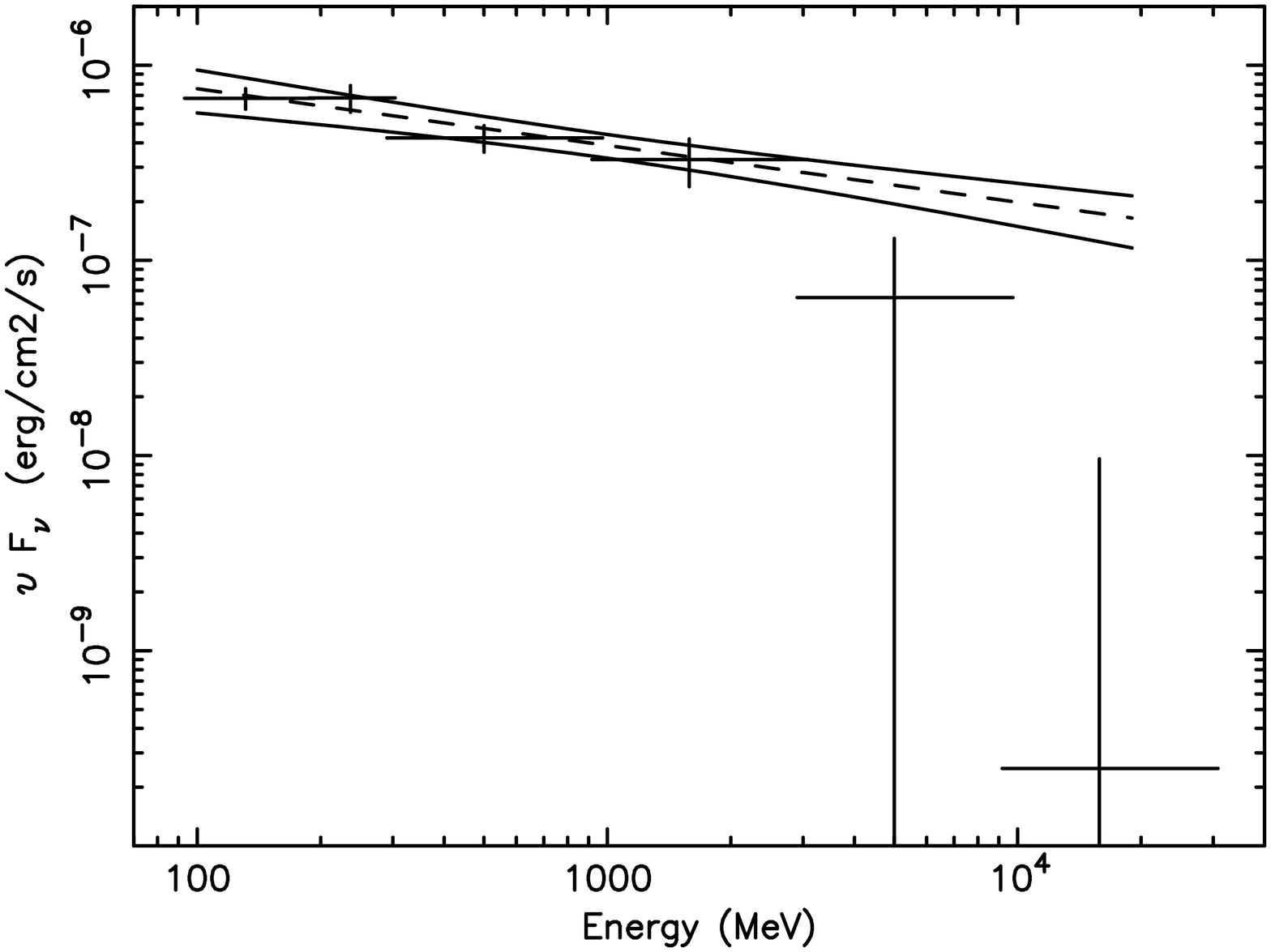}
  \caption{$\nu F_\nu$ spectrum of the data points from the LAT--only unbinned likelihood analysis of \GRB between \Tz{3.3} and \Tz{21.6}. Black dashed and solid lines show the best-fit power-law model and $\pm1~\sigma$ error contours, derived from the covariance matrix of the fit.}\label{Fig:latsed}
\end{figure}

In order to better characterize the power-law component at the highest energies,
we ran a LAT-only data analysis using the unbinned likelihood technique
for the full prompt phase. The fitted spectrum is shown in figure~\ref{Fig:latsed} (black points). The resulting photon index is $-2.29$ $\pm$ 0.09, much softer than the $-1.79\pm0.02$ index found for the joint GBM/LAT analysis. Considering the systematic effects in both analyses, this difference in photon index is significant ($\sim3\sigma$ level) and is an indication of the presence of a spectral break.
With the LAT data alone, we could not find any significant evidence for a deviation from the simple power-law shape, probably because of the limited lever arm in energy. 
Hence, we investigated this effect using the joint fits of the GBM and LAT data.

We fitted the GBM/LAT spectra with the combination of the Band function and a power-law model with an exponential cutoff (CUTPL),
\begin{equation}
\label{eq:cutoff}
 f(E) = B\left(\frac{E}{E_{\rm piv}}\right)^{\lambda}\exp\left(-\frac{E}{E_F}\right).
\end{equation}
Here $B$ is the normalization in units of photons s$^{-1}$ cm$^{-2}$ keV$^{-1}$, $E_{\rm piv}$ is the pivot energy fixed at 1~GeV, $E_F$ is the e-folding energy, and $\lambda$ 
is the power-law photon index.

\begin{figure}
  \begin{center}
  \rotatebox{0}{\includegraphics[width=0.9\linewidth]{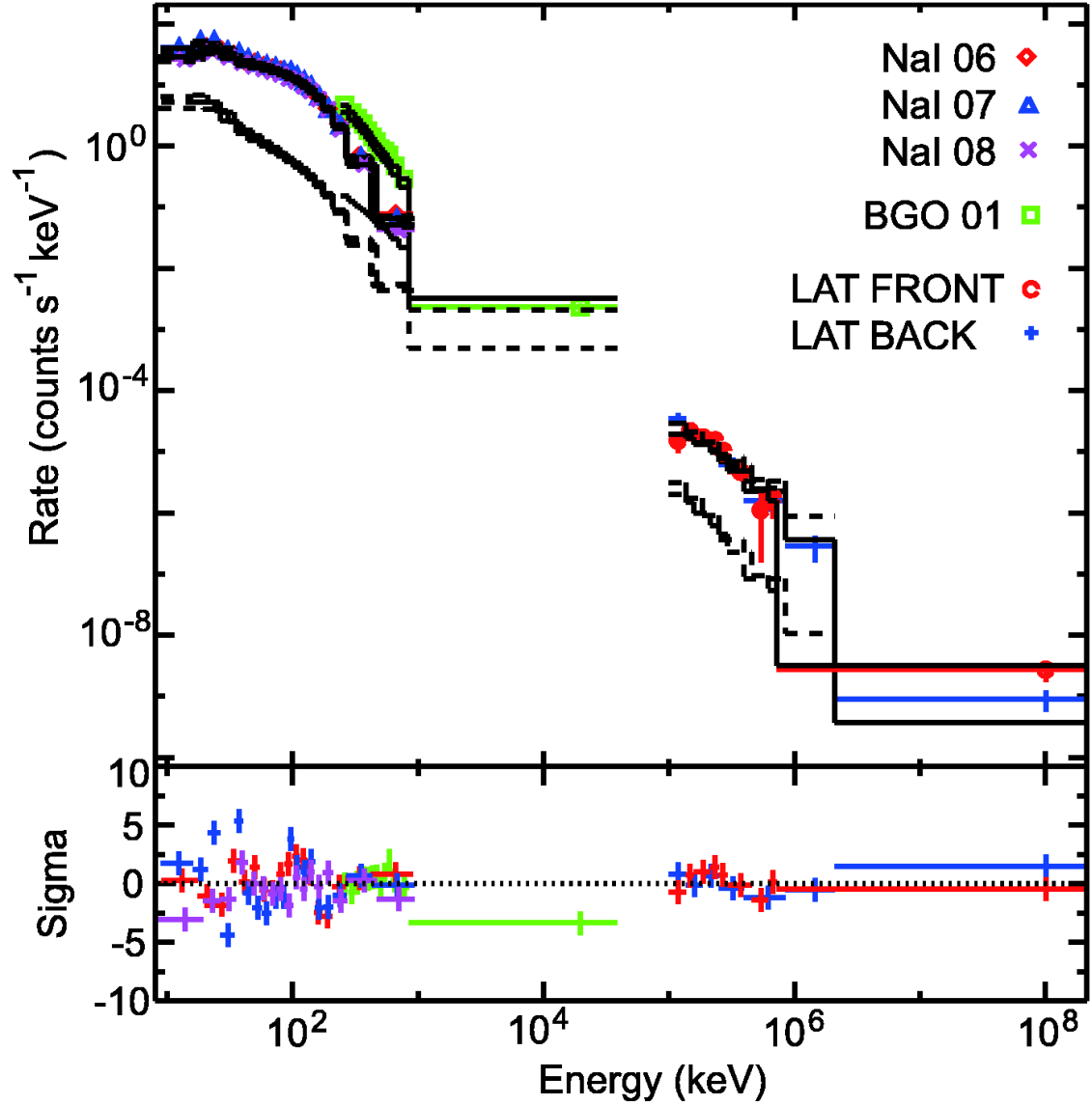}}
  %\rotatebox{-90}{\includegraphics[scale=0.4]{figure/Sigma_ba_all.pdf}}
  \caption{Joint spectral fitting of GBM and LAT data between \Tz{3.3}
and \Tz{21.6}. The top panel shows the count spectra and best-fit (Band+CUTPL)
 model (histograms). The lower panel shows the residual of the spectral fitting.  \label{Fig:spec}}
  \end{center}
\end{figure}

The fit results are summarized in table~\ref{tab:spec}, and the count spectra
and residuals are shown in figure~\ref{Fig:spec} for the best-fit model.
The e-folding energy is $E_F = 1.41_{-0.42}^{+0.22}\;stat.\,\pm0.30\;syst.$~GeV, while the power-law photon index below the cutoff energy is $\lambda\simeq-1.72_{-0.02}^{+0.10}\;stat.\,\pm0.01\;syst.$, which is a bit harder than in the (Band+PL) case.
The systematic uncertainties have been derived using the bracketing instrument response functions, as described in detail in~\cite{Fermi..GRB080825C}.
The parameters of the Band function change little from one fit to another. The C-STAT value for this model improves by 40.5 compared to the (Band+PL) model, which is significant at the $>4\sigma$ level (see the deeper discussion below).
%For completion, see ~\cite{baring06, Granot:08}, 
We also tried to fit the data with a broken power-law model,
\begin{equation}
\label{eq:bknpo}
f(E)~=~%
\left\{
\begin{array}{clc}
C\left(\frac{E}{E_{\rm piv}}\right)^{\lambda_l}& & {\rm for}\;E\leq E_{\rm break} \\
C\left(\frac{E_{\rm break}}{E_{\rm piv}}\right)^{\lambda_l}&\left(\frac{E}{E_{\rm break}}\right)^{\lambda_h} & {\rm for}\;E>E_{\rm break}
\end{array}
\right\},
\end{equation}
where $\lambda_l$ and $\lambda_h$ are the low- and high-energy power-law photon
indexes, respectively, $E_{\rm piv}$ is the pivot energy fixed at 1~GeV, and $E_{\rm break}$ is the
break energy.  However, the significance of the fit was close to that found using the (CUTPL) model so that we cannot distinguish between the two models. The fit with a broken power-law gave a break energy $E_{\rm break} = 219_{-56} ^{+65}$~MeV and a high-energy photon index of $\lambda_h=-2.47_{-0.17} ^{+0.14}$.

% Fixed as per Brigg's proposal
% Max Delta C-STAT is now 16.7
One may assess the significance of the spectral cutoff by computing the difference in the best-fit C-STAT values for the (Band+PL) and (Band+CUTPL) models.  Since C-STAT is equal to twice the log-likelihood, this is the standard likelihood ratio test; and conventionally, one calculates the significance of a change in log-likelihood using Wilks' theorem.  In this case, Wilks' theorem states the $\Delta$(C-STAT) values should be asymptotically distributed as $\chi^2$ for one degree of freedom.  However, certain assumptions are required for the validity of this calculation.  For the highest reliability, we studied the distribution of $\Delta$(C-STAT) values via simulations, creating $2 \times 10^4$ random realizations of the null hypothesis (the (Band+PL) model with parameters set at the best-fit values) and fit the data for each trial with both models.  In the resulting distribution of $\Delta$(C-STAT) values, the largest difference we found was 16.7, much smaller than the value of 40.5 for the actual data (see table ~\ref{tab:spec}). We therefore place a firm upper-limit on the probability that our fit of the exponential cutoff occurred by chance of $5 \times 10^{-5}$.  This corresponds to a Gaussian equivalent significance of 4.05$\sigma$.

Our distribution of $\Delta$(C-STAT) values shows a slight excess over the $\chi^2$ distribution at large values indicating that perhaps the asymptotic distribution has not been reached for this number of trials.  To be conservative, we do not evaluate the significance according to the conventional procedure of using the observed $\Delta$(C-STAT) value of 40.5 and the $\chi^2$ distribution.  Unfortunately, the number of simulations that would be required to determine the significance of the observed cutoff is prohibitive.  Nonetheless, the sizeable gap between the largest $\Delta$(C-STAT) value obtained in the simulations, 16.7, and the observed value of 40.5 suggests that the significance is much larger than 4$\sigma$.  For the 4 different sets of instrument response functions that we used in our study of the systematic uncertainties, we always found $\Delta$(C-STAT)$ \ge 32$.  The significance of the spectral cutoff will be hereafter quoted as $>4\sigma$.

% Numbers have been checked - 28/05/2010
%The parameters of the Band function change little from one fit to another.
Using the fit results for the best model (Band+CUTPL), 
we estimate a fluence of 2.07$\pm 0.04 \times 10^{-4}\,{\rm erg}\,{\rm cm}^{-2}$  (10~keV--10~GeV) from \Tz{3.3} to \Tz{21.6}.
%\Tz{3.328} to \Tz{21.632}.
These data give an isotropic energy
$\mathcal{E}_{\gamma, iso}$ = 2.24 $\pm 0.04 \times 10^{54}$~erg,
comparable to that of GRB~090902B \citep{Fermi...GRB090902B}.

\begin{figure}[htbp]
  \centering
  \rotatebox{0}{\includegraphics[scale=0.5]{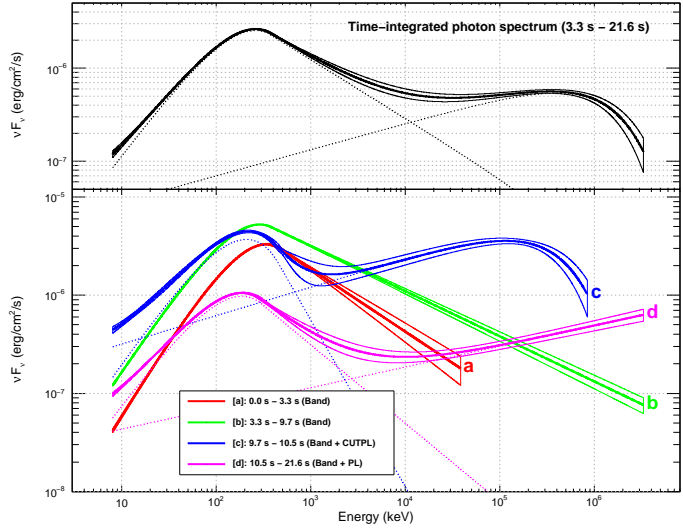}}
  \caption{{\em Top}: The best-fit (Band+CUTPL) model for the time-integrated data plotted as a $\nu F_\nu$ spectrum. The two components are plotted separately as the dashed lines, and the sum is plotted as the heavy line. The $\pm1~\sigma$ error contours derived from the errors on the fit parameters are also shown. 
 \label{Fig:TimeInt_uFu}
 {\em Bottom}: The $\nu F_\nu$ model spectra (and $\pm1~\sigma$ error contours) plotted for each of the time bins considered in the time-resolved spectroscopy.
  \label{Fig:stacked_uFu}
}
\end{figure}

%\subsection{Time Resolved Analysis and the 10~s peak}
%\label{section:correlation}
%\label{section:10speak}
We then performed a time-resolved spectral
analysis of the prompt phase in the four time intervals \emph{a, b, c, d}. The spectra are shown in figure \ref{Fig:stacked_uFu}, and the results  are summarized in table \ref{tab:extra}, where the best-fit parameters are given for the statistically preferred model, and the C-STAT values are given for the various models.
The extra power-law component is found to be very significant in intervals \emph{c} and \emph{d}, but not at the beginning of the prompt phase in intervals \emph{a} and \emph{b}. The spectral cutoff is significant at the $>4\sigma$ level only in the common sharp peak (time interval \emph{c}), where the GeV flux is the highest, but is only marginally significant ($\sim4\sigma$) in time bin \emph{d}.

In time interval \emph{b}, the improvement in the fit statistics when adding the extra power-law component is only $\Delta$(C-STAT) = 11.6.  As a consequence, the parameters of the power-law are not very well constrained, yielding a normalization $B=2.9^{+6.4}_{-1.0}$ $10^{-10}\;$photons cm$^{-2}$ s$^{-1}$ keV$^{-1}$ and a power-law index $\lambda=1.7^{+0.2}_{-0.1}$.
In time interval \emph{c}, we found the cutoff energy to be $E_F = 0.40_{-0.06}^{+0.13}\;stat.\,\pm0.05\;syst.$~GeV (table \ref{tab:extra}).
Note that we fixed the pivot energy  at $E_{\rm piv} = 1$~MeV for
time interval \emph{c}, since this is the only interval where the extra power-law component is dominant over the Band component at very low energies, and setting $E_{\rm piv} = 1\;$GeV resulted in very asymmetric
and very large uncertainties, especially for the normalization $B$ of the extra power-law component.
We also tried to fit time interval \emph{c} with a broken power-law model, see equation~\ref{eq:bknpo}; but again the fit significance was close to that of the (Band+CUTPL) model, so that we cannot distinguish between the two models.  The fit with a broken power law gave
a break energy $E_{\rm break} = 264^{+233}_{-75}$~MeV and a photon index above $E_{\rm break}$ of $\lambda_{h} = -3.55^{+0.63}_{-3.28}$.
In time interval \emph{d}, the improvement in the fit statistics when adding a cutoff to the the extra power-law component is only 17.4 (roughly $\sim4\sigma$), which is quite high, but not sufficient to claim the presence of an energy cutoff in this bin alone.  However, as the cutoff is strong in the preceding time interval \emph{c}, we looked at the behavior of the e-folding energy. For time interval \emph{d}, the e-folding energy  is found to be $E_F = 2.21^{+0.92}_{-0.69}$~GeV, which is much higher than the one found in interval \emph{c} (the 2-$\sigma$ confidence intervals for the cutoff in bins \emph{c} and \emph{d} actually exclude each other). This indicates a possible time evolution of the high energy cutoff.
%power-law normalization $B=4.9^{+1.4}_{-0.9}$ $10^{-10}\;\gamma$ cm$^{-2}$ s$^{-1}$ keV$^{-1}$
%power-law index $\lambda=1.71^{+0.14}_{-0.04}$.
%bin c: E=400MeV -174 + 395, and bin d: E=2110MeV -1130 +3480
 
\subsection{LAT extended emission}\label{section:extended}
As the burst was occulted by the Earth from \Tz{540} to \Tz{3000}, we performed the unbinned likelihood analysis using `transient' class events in the time interval from \Tz{20} to \Tz{300} (a small margin is needed to safely define a circular ROI), and use `diffuse' class events after \Tz{3000}~\footnote[1]{See \cite{2009ApJ...697.1071A} for the definitions and recommended usage of the LAT event classes.}.
`Transient' class events are treated as in \S\ref{section:rmfit}. In addition, for the `diffuse' class events, we included in the model the standard galactic background component, described by the FITS model file {\tt gll\_iem\_v02.fit}, with fixed normalization, and the standard isotropic background component, whose spectrum is given in the model file {\tt isotropic\_iem\_v02.txt}, with the normalization left free.  Both model files may be downloaded from the FSSC website.

We divided the LAT data into several time intervals, using intervals \emph{a,b,c,d} for the prompt phase, and modeled the GRB extended emission spectrum as a power-law. For the period \Tz{3000} -- \Tz{4800}, the fit resulted in a test statistic of 29.4, 
corresponding to a detection at a $\sim5\sigma$ level, which is remarkable for a time period $\sim1$ hour after the burst.

\begin{figure}[htbp]
  \centering
  \includegraphics[scale=0.7, angle=0]{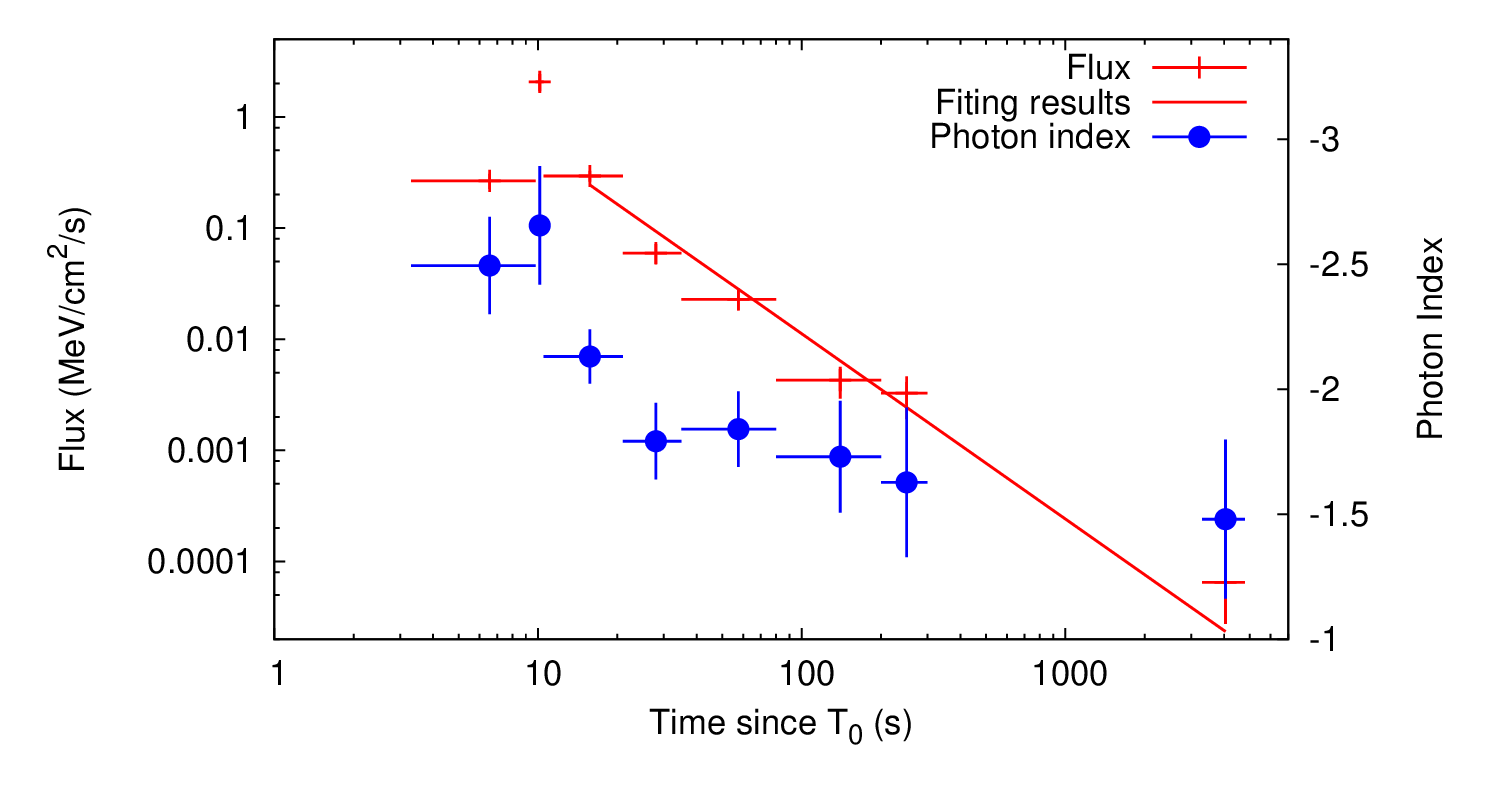}
  \caption{Time variation of the LAT flux (red cross) and photon index (blue filled circle) 
  for the extended emission of \GRB. After the end of the prompt emission at \Tz{21.6}, the flux decays following a power-law with index $-1.69\pm0.03$ (red solid line).}
  \label{Fig:extended}
\end{figure}

Figure \ref{Fig:extended} shows the flux and photon index versus time.
The LAT flux follows a power-law with time-dependence $(T-T_0)^{-1.69\pm0.03}$ after \Tz{21.6}, 
similar to the behavior of bursts GRB~090510 and GRB~090902B 
\citep{Fermi...GRB090902B, Nature..GRB090510, 2010ApJGRB090510}.
Prior to \Tz{21.6}, the photon index varies significantly with values ranging from $-2.5$ to $-1.7$.  By contrast, after \Tz{21.6},
the photon index is almost constant with values in the range $-1.5$ to $-1.9$.  The soft spectral index in time interval \emph{c} is consistent with the spectral break of the extra component described in section~\ref{section:rmfit}, and the gradual hardening from time bin \emph{d} is consistent with its disappearance.

\begin{table}
\begin{small}
\caption{Summary of GBM/LAT joint spectral fitting between \Tz{3.3} and 
\Tz{21.6}. The flux range covered by both instruments is 10~keV--10~GeV.}
\label{tab:spec}
\begin{center}
\begin{tabular}{p{5cm}cccc}
\hline \hline
Fitting model & Band  & Band+PL &  Band+CUTPL \\ \hline \hline
\multicolumn{4}{l}{\it Band function}\\
A ($\gamma$ cm$^{-2}$ s$^{-1}$ keV$^{-1}$) 
                      & 0.176 $\pm$ 0.002 & 0.173 $\pm$ 0.003   &0.170   $_{-0.004}^{+0.001}$ \\
$E_{\rm{peak}}$ (keV) &   249 $\pm$ 3      & 256 $\pm$   4      & 259  $_{-2}^{+8}$     \\
$\alpha$ (index 1)    & $-0.71 \pm$ 0.01   & $-0.62\pm$ 0.03   & $-0.64_{-0.09}^{+0.02}$  \\
$\beta$ (index 2)     & $-2.30 \pm$ 0.01   &$-2.59_{-0.05}^{+0.04}$& $-2.63_{-0.12}^{+0.02}$  \\\hline

\multicolumn{4}{l}{\it Power-law}\\
B ($10^{-10}\;\gamma$ cm$^{-2}$ s$^{-1}$ keV$^{-1}$)&-&3.17 $_{-0.33}^{+0.35}$&5.80  $_{-0.60}^{+0.81}$ \\
$\lambda$ (index)                                           &-& $-1.79\pm$ 0.02        & $-1.72_{-0.02}^{+0.10}$ \\ 
$E_{\rm piv}$                                               &-& 1~GeV (fixed)           & 1~GeV (fixed) \\ \hline
\multicolumn{4}{l}{\it High-energy cutoff}\\
$E_F$ (GeV)                    & -                 & -                    & 1.41 $_{-0.42}^{+0.22}$   \\ \hline

%\multicolumn{4}{l}{\it Effective area correction}\\
%BGO/LAT (back)   &  0.79 (fixed)  &  0.79 (fixed)  &   0.79 (fixed)    \\
\hline
Flux ($\gamma$ cm$^{-2}$ s$^{-1}$)      &  42.2$\pm$0.1  &  43.5$\pm$0.3 &  43.3$\pm$0.2  \\
Flux ($10^{-5}$ erg~cm$^{-2}$ s$^{-1}$) &  1.18$\pm$0.01 &  1.15$\pm$0.02 & 1.13$\pm$0.02  \\
\hline \hline
C-STAT / DOF     &  1395.1 / 579 & 1287.8 / 577 & 1247.3 / 576 \\
$\Delta$(C-STAT)$\dag$ &       -      & 107.3        & 40.5         \\
\hline \hline
\multicolumn{4}{l}{\footnotesize{$\dag$ \it with respect to the preceding model (column).}}\\
\end{tabular}
\end{center}
\end{small}
\end{table}

%\end{landscape}

%\begin{landscape}

\begin{table}
\begin{footnotesize}
\caption{Summary of GBM/LAT joint spectral fitting by best 
 model in 4 time intervals. The flux range covered by both instruments is 10~keV--10~GeV.} 
\label{tab:extra}
\begin{center}
\begin{tabular}{p{5cm}cccc}
\hline \hline
Time interval from $T_0$ (s)& (a) 0.0--3.3 & (b) 3.3--9.8 & (c) 9.8--10.5          & (d) 10.5--21.6 \\ \hline
Prefered Model                  & Band         & Band         & Band+CUTPL             & Band+PL        \\ \hline \hline
\multicolumn{5}{l}{\it Band function}\\
A ($\gamma$ cm$^{-2}$ s$^{-1}$ keV$^{-1}$)
                   &0.146 $\pm$ 0.004   &0.302  $\pm$ 0.004 &0.335 $_{-0.012}^{+0.064}$ &0.100 $\pm$ 0.005  \\
$E_{\rm{peak}}$    &338   $\pm$ 10      & 288   $\pm$ 4     & 209 $_{-16}^{+5}$         & 186  $\pm$ 6    \\
$\alpha$ (index 1) &$-0.42 \pm 0.03$  & $-0.55 \pm 0.01$   &$-0.59_{-0.06}^{+0.39}$   &$-0.70^{+0.07}_{-0.06}$\\
$\beta$ (index 2)  &$-2.64_{-0.09}^{+0.07}$ & $-2.46 \pm$ 0.02  & $-3.69_{-0.53}^{+1.81}$   & $-2.80_{-0.18}^{+0.13}$  \\\hline
\multicolumn{5}{l}{\it Power-law}\\
B ($10^{-10}\;\gamma$ cm$^{-2}$ s$^{-1}$ keV$^{-1}$)
                               &-& -	             & 7.56 $^{\dag}$ $_{-0.50}^{+2.25}$ 	& 3.07 $^{+0.38}_{-0.36}$	\\
 $\lambda$ (index)                                         
                               &-& -                 & $-1.71^{+0.02}_{-0.05}$	        & $-1.79 \pm 0.03$	\\
$E_{\rm piv}$                  &-& -                 & 1~MeV (fixed)	                & 1~GeV (fixed)	\\ \hline
\multicolumn{5}{l}{\it High-energy cutoff}\\
$E_F$ (GeV)                    &-& -	             & 0.40 $_{-0.06} ^{+0.13}$ & -	\\ \hline

%\multicolumn{5}{l}{\it Effective area correction}\\
%BGO/LAT (back)&	0.79 (fixed) &	0.79 (fixed)  &	0.79 (fixed)&	0.79 (fixed)		\\
\hline

Flux ($\gamma$ cm$^{-2}$ s$^{-1}$)     &31.4 $\pm$0.2  & 66.4 $\pm$ 0.3  & 109.0 $\pm$1.4	       &25.5 $\pm$0.2	\\
Flux ($10^{-6}$ erg~cm$^{-2}$ s$^{-1}$)&9.96 $\pm$0.41 & 18.9 $\pm$0.30 & 29.22 $\pm$1.60	       & 5.83 $\pm$0.30 \\
\hline \hline
\multicolumn{5}{l}{\it C-STAT / DOF  }\\
Band               & 622.4 / 579 & 944.2 / 579 & 655.9 / 579 & 1033.8 / 579 \\
Band+PL            & 624.3 / 577 & 932.6 / 577 & 598.7 / 577 & 950.6 / 577 \\
Band+CUTPL         & 618.8 / 576 & 928.3 / 576 & 574.2 / 576 & 933.2 / 576 \\
\hline
\multicolumn{5}{l}{\it $\Delta$(C-STAT)}\\
Band$\rightarrow$(Band+PL)        & -1.9     & 11.6        & 57.2        & 83.2\\
(Band+PL)$\rightarrow$(Band+CUTPL)&  5.5     & 4.3         & 24.5        & 17.4\\
\hline \hline  
\multicolumn{5}{l}{\it\footnotesize{$\dag$ As ${\rm E}_{\rm piv}$ {\rm = 1~MeV}, {\rm B} has the unit of $10^{-4}\;\gamma$ cm$^{-2}$ s$^{-1}$ keV$^{-1}$}} 
\end{tabular}
\end{center}
\end{footnotesize}
\end{table}

\clearpage
%\documentclass[12pt,preprint]{aastex}
%\documentclass{emulateapj}
%\usepackage{apjfonts}
%\usepackage{amsmath}
%
%
%\newcommand{\Tz}[1]{{$T_0+#1$\,s}}

%\newcommand{\fpm}[1]{{\bf #1}}

%\begin{document}

\section{Discussion and Interpretation}

\subsection{Prompt Emission Phase}
\label{sec:discussion}
%The {\it Fermi} observation of GRB~090926A shows that 
%the spectrum for the whole time interval of the prompt emission (between
%\Tz{3.3} and \Tz{21.6}) can be well fitted by the combination of
%the canonical Band function component peaking at $\simeq 250\;$keV
%and an extra power-law component dominant above $\simeq 10\;$MeV.
%For the time-resolved spectra, the time intervals \emph{c}
%(between \Tz{9.8} and \Tz{10.5}) and \emph{d} (between \Tz{10.5} and \Tz{21.6}) 
%clearly have an extra power-law component in addition
%to the Band component.
%%%%%%%%%%%%%%%
The {\it Fermi} observations show that GRB~090926A clearly has an extra high-energy 
component in addition to the Band component in the time-integrated as well as in the time-resolved spectra.
%%%%%%%%%%%%%%%
This is the third case of a LAT detection of an
extra spectral component, after GRB~090510 and GRB~090902B
\citep{2010ApJGRB090510, Fermi...GRB090902B}.
That is, we have such detections in 3 out of the 4 brightest LAT GRBs
\citep[except GRB 080916C, see][]{2009Sci...323.1688A}.
Since we require a confidence level of $>5 \sigma$ to claim a
detection, we can unambiguously identify it only in the brightest LAT GRBs, which
suggests that such a component may be intrinsically very common in GRBs.

%Our best fit spectral models for the intervals \emph{c} and \emph{d},
%where the extra power-law components have been for simplicity assumed to extend to
%the lowest energy range, show that the extra power-law components are clearly dominant 
%at $\gtrsim 1\;$MeV for \emph{c} and at $\gtrsim 10 \;$MeV for \emph{d}, while 
%the deviations from the simple Band components due to the extra emission 
%at $\lesssim 20\;$keV are less significant for both time intervals. 

The behavior in time interval \emph{c} is remarkable, as the light curve shows a
clear spike-like structure in the LAT energy range that is dominated by the extra spectral component. 
%%%%%%%%%%%%%%%%%%
The spike in all the energy ranges has the peak times synchronized within
50~ms and shows similar pulse widths
(see Figure~\ref{Fig:prompt light curves zoom}), which indicates that the origins
of the lowest and highest energy emission components as well as the Band component 
are related,
%%%%%%%%%%%%%%%%%%
%The light curve above 100 MeV is highly correlated with that of the
%energy range of $8-14.3\;$keV (see Table~1), which indicates a
%common origin for the lowest and highest energy emission components
i.e., they could either arise from the same physical region and
possibly also the same spectral component, or are otherwise directly
physically linked, such as photons generated in one emission region
being scattered by electrons in the other emission region 
(see specific models discussed below).
%Furthermore, the light curves in the energy ranges of 14.3 - 260 keV 
%and of 260 keV - 5 MeV are marginally correlated with those in the 
%lowest and highest energy bands (see Table~\ref{tab:corr}), and the peak times 
%of the brightest pulses are synchronized within 100 ms with similar 
%pulse durations in all the energy bands (see Figure~\ref{Fig:prompt light curves zoom}). These imply 
%that the origins of the Band component and the extra power-law 
%component (or the lowest and highest energy emission components) are 
%also related in the time interval \emph{c}.

The delayed onset of the LAT emission 
is common to almost all the LAT GRBs \citep[except GRB~090217A, see][]{2010ApJGRB090217}.
The delay may arise from the following four effects: (1) a
flux increase of the Band component, (2) a hardening of the Band
component (i.e., increase of the peak energy $E_{\rm{peak}}$ and/or high-energy
spectral index $\beta$), (3) a flux increase of
the extra component, or (4) an increase of the cutoff energy in the spectrum.
However, effect (4) does not seem to be a major effect for LAT
GRBs so far since there is no clear sign of a high-energy cutoff or
steepening in the spectra before the LAT onsets \citep[see also the
discussion on GRB~080825C in][]{Fermi..GRB080825C}. 
The LAT detection of GRB~090926A starts from time interval \emph{b}, and the clear emergence of the extra component occurs even later, $\approx 10\;$s after the onset of the Band component, so that this delay
is likely to be due to the combination of (1) and (2).
However, while no significant extra component is detected in time intervals \emph{a} and
\emph{b}, it may still be present with a lower cutoff energy that falls under or close 
to the Band function model component, so that we may not exclude contributions
from effects (3) or (4) to the delayed onset.

There are several theoretical models for the origin of the extra
spectral component.  The delayed extra component could be emitted
from a forward shock that propagates into the external medium
\citep{1997ApJ...476..232M,1998ApJ...497L..17S}, while the Band
component is thought to have a separate origin.
The delay timescale of the extra spectral component would correspond to
the time needed for the forward shock to sweep up material and
brighten \citep{2009MNRAS.400L..75K,ghisellini09,razzaque10b}.
%While the radius of an emitting shell in an internal emission
%model is $R \lesssim \Gamma^2 c \Delta T/(1+z)$, where $\Gamma$ is the
%bulk Lorentz factor and $\Delta T$ is the pulse duration, the forward
%shock radius in an external shock model is $R_f \approx \Gamma_f^2 c
%T_{\rm dur}/(1+z)$, where $\Gamma_f$ is the Lorentz factor of the
%forward shock and $T_{\rm dur}$ is the duration of the burst.
The rapid variability observed in GRB~090926A is contrary to
expectations from an external shock model, unless it is produced by
emission from a small portion of the blast wave within the Doppler
beaming cone. This could occur, for instance, if the external medium
is clumpy on length scale $\approx \Gamma_f c \Delta T/(1+z) \simeq
10^{12}\; (\Gamma_f/10^3) (\Delta T/0.2\;{\rm s})\;$cm,
%%%%%%%%%%%%%%%%
where $\Gamma_f$ is the Lorentz factor of the forward shock and $\Delta T$
is the pulse duration 
%%%%%%%%%%%%%%%%
\citep{1999ApJ...513L...5D,dermer08}. 
This is based on interactions between a very thin shell, 
%(with a width $\ll R_f/\Gamma_f^2$), 
prior to the onset 
of the self-similar expansion phase, and an external medium with very small scale clumps.
If the extra component is synchrotron emission from the
forward shock, then the synchronization of the pulse peak times of the
Band and extra component requires an explanation.  One possibility is
that the extra component arises from inverse Compton (IC) scattering of
the radiation of the Band component by the high-energy electrons in
the forward shock.

As for internal emission models, in which both 
spectral components arise within the ejecta, the extra
component can be produced by IC scattering by energetic leptons or
via hadronic processes. In either case, the time of the peak of the extra
component would lag relative to the Band component in the same emission
episode, although the time lag can be limited by a timescale
comparable to the pulse duration, which would still be consistent with the
observed synchronization of the two components.  A simple
leptonic model could comprise synchrotron plus synchrotron
self-Compton (SSC) emission. 
Under the physical conditions typically assumed in the internal shock model \citep{rees94},
all electrons emitting synchrotron emission cool on a timescale much shorter
than the dynamical time (i.e., the electrons are in the fast cooling regime), so
that the photon index $\alpha$ below $E_{\rm{peak}}$ should be $-1.5$, which is not consistent
with the results from our fits, $-0.7 \lesssim \alpha \lesssim -0.4$. The synchrotron plus
SSC model would need to overcome this problem.
%The extra component could be produced by
%the IC of an external photon field \citep[e.g., a delayed X-ray cocoon
%emission;][]{toma09}. This scenario predicts different
%temporal behaviors of the seed X-ray photons and the up-scattered
%high-energy photons, however, which are not consistent with the observations of
%this burst.  The external photon field could be the variable
%quasi-thermal emission from the photosphere of the jet, i.e., the Band
%component is the photospheric emission and the extra component is the
%IC of the photospheric emission by electrons in the dissipation region
%at large radius
In the photospheric emission model of the Band component \citep[e.g.,][]{2000ApJ...530..292M},
the extra component could be the IC of the photospheric emission by electrons in the 
dissipation region at large radius 
\citep{beloborodov09,gao09,ryde09}.
In this model, the low energy excess seen in interval \emph{c} could be synchrotron emission of the electrons and
the electron-positron pairs created by the cascade process, and the delay timescale 
of the extra component could 
be explained by the evolution of the jet physical conditions \citep{toma10}.

Hadronic processes, such as a photopion-induced pair cascade or proton/ion
synchrotron emission \citep{2009ApJ...705L.191A,2009arXiv0908.0513R,Wang:09b}
can make a spectral component that is distinct from that which is
commonly observed during the prompt phase. Large values of the bulk Lorentz factor of the emitting region, $\Gamma \gtrsim 10^3$, imply large energy requirements for significant high-energy emission in either photo-hadronic or proton/ion synchrotron
models. The bulk Lorentz factor that is inferred by the argument of 
the pair absorption opacity for this burst, however, is relatively low 
($\Gamma \sim 200$--$700$, see below).
Thus for proton/ion synchrotron models, which require a heavily magnetically
loaded shocked jet, the total energy requirements $\propto \Gamma^{16/3}$ 
\citep{Wang:09b,2009arXiv0908.0513R}
are smaller by a factor of $\sim 7$--$5000$ compared to the case of $\Gamma \sim 10^3$,
much improving the viability of such models.
The lower allowed values of $\Gamma$ also reduce the energy requirements in photo-hadronic models, where non-thermal protons usually dominate the bulk energy~\citep{2009ApJ...705L.191A,Wang:09b,2010ApJGRB090510}. Under the assumption that the photon field is homogeneous and steady in the emitting region, the efficiency of photo-hadronic interactions by high-energy protons at the peak of the Band spectrum component in this burst is estimated to be a few percent for $\Gamma \sim 10^3$. This efficiency scales as $\Gamma^{-4}$, which implies significant reduction of the total energy requirements for the lower allowed values of $\Gamma$. However, $\Gamma \lesssim 700$ is indicated under the assumption that the photon field is inhomogeneous and time-dependent, as will be discussed below, for which it is not so clear how the photo-hadronic interaction efficiency (as well as the neutrino production efficiency, see~\cite{murase06,dermer07,razzaque09}) depends on $\Gamma$.

Another remarkable aspect of this burst is the spectral break (or cutoff)
of the extra component that has been measured in the time-integrated spectrum
of the prompt emission and for time interval \emph{c} with a high significance 
($>4\,\sigma$; see sections \ref{section:rmfit}, 
%\ref{section:correlation} 
and tables \ref{tab:spec}, \ref{tab:extra}).
This cutoff may be due to pair production
($\gamma\gamma \to e^+e^-$) within the emitting region, although we
cannot rule out the possibility that there is an intrinsic spectral
break related to the energy distribution of the emitting particles or
the emission mechanism (e.g., IC scattering in the Klein-Nishina regime).  
Absorption by the extragalactic background light (EBL) cannot cause
this spectral feature since the opacity at the observed break energy
for the redshift of GRB~090926A is very small for practically all EBL models
\citep[][and references therein]{2009arXiv0905.1115F}.
%\citep{2003A&A...407..791M, kneiske04,2006ApJ...648..774S,
%2008A&A...487..837F,gilmore09,2009arXiv0905.1115F}. 
We focus on the spectral feature in time interval \emph{c} to constrain
the physical properties of the emitting region by
introducing the critical photon energy $E_c$ at which the pair
production opacity is unity, $\tau_{\gamma\gamma}(E_c) = 1$
\citep[e.g.,][]{krolik91,fenimore93,2001ApJ...555..540L}.  

In order to characterize the spectral break, we have fit the data with a model that consists of
an extra power-law component modified by absorption due to pair
production, but this spectral model is not
unique. Although the instantaneous emission from a thin shell exhibits
a photon spectrum like $f \propto E^{\lambda} \exp
(-\tau_{\gamma\gamma}(E))$, the shape of the time-integrated spectrum of
a single pulse may depend on the details of the emission mechanism
\citep{baring06,Granot:08}.  For example, the simple model of an
emitting slab leads to $f \propto E^{\lambda} [1-
\exp(-\tau_{\gamma\gamma}(E))]/\tau_{\gamma\gamma}(E)$, which is a
smoothly broken power-law spectrum since $\tau_{\gamma\gamma}(E)$ is a
power-law function of $E$ when the intrinsic emission spectrum is a
power-law function (see below). 
A fully time-dependent and self-consistent semi-analytic
calculation featuring emission from a very thin spherical shell over a
finite range of radii \citep{Granot:08} would also lead to a smooth break to a
steeper power-law in the time-integrated spectrum of a single pulse.
%in the time integrated (over a single spike in the
%light curve) spectrum -- see more detailed comments on this model below. 
%Thus, the shape of the spectral turnover (or break) may
%serve as a distinctive signature of intrinsic pair attenuation,
%therein affording powerful environmental diagnostics \citep[e.g.,][]{baring06}.
%As an example, an even more powerful signature is expected in the time-dependent model of
%\citet{Granot:08}, where photons above the spectral break energy are 
%expected to arrive predominantly near the onset of the spike. 
%These expected signatures may afford a more unambiguous indicator of the
%intrinsic opacity to pair production
%when high photon counts permit extension of the spectrum well above the observed break.
%Such an opportunity is
%not available for GRB~090926A, however, since there are not enough
%photons above the spectral break during the sharp spike in time
%interval \emph{c}. 
In section~4.2, we have fitted the observed extra spectral component for time interval 
\emph{c} by two empirical functions: a power-law with exponential cutoff (Eq.~1) and 
a broken power-law model (Eq.~2). 
%For the latter model, we have
%obtained the photon index above the break energy 
%$\lambda_h \simeq -3.55^{+0.63}_{-3.28}$. 
However, the photon counts are not enough to distinguish between the two models.
%%%%%%%%%%%
In the time-dependent model of \citet{Granot:08}, photons above the spectral break energy 
are expected to arrive predominantly near the onset of the spike. This signature may 
afford a more unambiguous indicator of the intrinsic opacity to pair production.
Such an opportunity is also unavailable for GRB 090926A, however, due to insufficient photon counts above the spectral break.
%%%%%%%%%%%
Here we only use the result of the
former model (Eq.~1), and consider the e-folding energy $E_F$ from the fit to
be good approximation of $E_c$.

In order to derive the pair absorption function
$\tau_{\gamma\gamma}(E)$, we first consider a simple model in which the
photon field in the emitting region is uniform, isotropic, and
time-independent in the comoving frame \citep[see the supporting
material for][]{2009Sci...323.1688A}.
We assume that the opacity at the photon energy around $E_c$ is
dominated by the extra power-law component itself, instead of the Band
component. This assumption is justified for the observed spectrum in
time interval \emph{c}, as shown below.  Let us define the observed
photon number spectrum of the extra component for one pulse, below the
break energy, as $f(E) = f(E_{\rm piv}) (E/E_{\rm piv})^{\lambda}$ in
units of ${\rm photons}\;{\rm cm}^{-2}\;{\rm keV}^{-1}$.  The energy
distribution of the photons in the comoving frame of the emitting
region is written as
\begin{equation}
n'_{\gamma}(E') = \left(\frac{d_L}{R}\right)^2 \frac{\Gamma f(E_{\rm piv})}{(1+z)^3 W'} 
\left(\frac{E'}{E'_{\rm piv}}\right)^{\lambda},
\end{equation}
where the quantities with a prime are measured in the comoving frame,
%$W'$ is the comoving radial width of the emitting region, and
$d_L \simeq 5.17 \times 10^{28}\;$cm is the luminosity distance of the source,
%%%%%%%%%%%%%%%%%
and $R$, $\Gamma$, and $W'$ are the distance from the central engine, the bulk
Lorentz factor, and the comoving radial width of the emitting region, respectively.
%%%%%%%%%%%%%%%%%
Photons with energy $E' = E'_c > m_e c^2$ annihilate mainly with target
photons with energy $E'_{\rm ann} \sim 2 m_e^2 c^4/E'_c$.  Then the
optical depth is of the order of $\tau_{\gamma\gamma}(E'_c) \sim 0.1
\sigma_T E'_{\rm ann} n'_{\gamma}(E'_{\rm ann}) W'$, where the pair
production cross section is approximated to be 0.1 times the Thomson
cross section $\sigma_T$.  More accurately, we have
\begin{equation}
\tau_{\gamma\gamma}(E'_c) = 
\sigma_T \left(\frac{d_L}{R}\right)^2 \frac{\Gamma E'_{\rm piv} f(E_{\rm piv})}{(1+z)^3} 
\left(\frac{E'_c E'_{\rm piv}}{m_e^2 c^4}\right)^{-\lambda-1} F(\lambda) = 1
\end{equation}
\citep{2009Sci...323.1688A}
where $F(\lambda) \approx 0.597 (-\lambda)^{-2.30}$ for $-2.9 \leq
\lambda \leq -1.0$.  The relation $R \simeq \Gamma^2 c \Delta T/(1+z)$
is valid for a large class of emission mechanisms, where $\Delta T$ is
the variability time. Then we obtain
\begin{equation}
\Gamma \simeq \left[\sigma_T \left(\frac{d_L}{c \Delta T}\right)^2 E_{\rm piv} f(E_{\rm piv}) F(\lambda)
(1+z)^{-2(\lambda+1)} \left(\frac{E_c E_{\rm piv}}{m_e^2 c^4}\right)^{-\lambda-1}\right]^
{\frac{1}{2(1-\lambda)}}.
\end{equation}
We can estimate $\Gamma$ and $R$ from the values of $E_{\rm piv}$,
$f(E_{\rm piv})$, $\lambda$, and $E_c$, which are inferred from the
observed spectrum.  We have estimated the variability timescale of the
Band component to be $\Delta T \simeq 0.15 \pm 0.01\;$s from the analysis
of the GBM emission in section~3.  From the
synchronization of the peak times and the similar pulse durations in
all energy ranges, we may assume that the variability timescale of the
extra component is the same.  The power-law with exponential cutoff
model with $E_{\rm piv} = 1\;$MeV results in 
$B = 7.56^{+2.25}_{-0.50} \times 10^{-4} \; {\rm photons}\;{\rm cm}^{-2}\;{\rm keV}^{-1}$, 
$\lambda = -1.71^{+0.02}_{-0.05}$,
and $E_c = E_F = 400^{+130}_{-60}\;$MeV. We may estimate the normalization of
the spectral fluence over the variability time $\Delta T$
around the spike by $f(E_{\rm piv}) \simeq 2 B \Delta T$, where $B$ is
the normalization of the time-averaged spectral flux over interval \emph{c}.
Then we obtain $\Gamma \simeq 720 \pm 76$, where
the error on $\Gamma$ has been calculated by the statistical errors of the 
parameters $\lambda, E_F,$ and $\Delta T$ as well as the error of $f(E_{\rm piv})$ taken as $\pm B \Delta T/2$.
This error on $\Gamma$ is much smaller than the uncertainty
of $\Gamma$ due to modeling the space inhomogeneity and the time dependence
of the target photon field discussed below.
The energy of the main target photons for the photons with $E_c$ is $E_{\rm ann} \sim 
2 \Gamma^2 m_e^2 c^4/[(1+z)^2 E_c] \simeq 70 \;(\Gamma/700)^2 (E_c/400\;{\rm MeV})^{-1}\;{\rm MeV}$,
while the extra power-law component is dominant above $\sim 1\;$MeV. Thus, our assumption 
that the target photons for the photons with energy $E_c$ are from the extra component is justified.
In this model, the emission radius is estimated to be 
$R \simeq \Gamma^2 c \Delta T/(1+z) \simeq 7 \times 10^{14}\;(\Gamma/700)^2 (\Delta T/0.15\;{\rm s})\;$cm.

A fully time-dependent and self-consistent semi-analytic model by \citet{Granot:08}
results in a significant reduction in $\tau_{\gamma\gamma}$ and
in the inferred value of $\Gamma$ 
%(or of the lower limit $\Gamma_{\rm
%min}$, if there is only a lower limit on a possible break energy) 
by a factor of $\sim 3$ compared to simpler models like the above
calculation, i.e., $\Gamma \simeq 220$ for this burst. 
Under the typical physical conditions for the shock emitting the bright $\gamma$-rays,
electrons are in the fast cooling regime, so that most of the radiation is emitted 
within a very thin layer behind the expanding shock front \citep[e.g.,][]{2000ApJ...534L.163G}.
The reduction in $\tau_{\gamma\gamma}$ occurs mainly since the high-energy photons
are emitted from a very thin cooling layer, 
so that those that are emitted from angles $<1/\Gamma$ relative
to the line of sight immediately propagate ahead of the shock front,
and can therefore potentially pair produce only with photons that
propagate at fairly small angles, $\theta_{12}$, relative to their own
direction.  The small values of $\theta_{12}$ suppress the interaction
rate ($\tau_{\gamma\gamma} \propto 1-\cos\theta_{12}$), and increase
the threshold energy for pair production, $E_1 E_2 (1+z)^2 > 2(m_e
c^2)^2/(1-\cos\theta_{12})$, where $E_1$ and $E_2$ are measured at
Earth. The time dependence also reduces the
time-averaged opacity over a single spike in the light curve, since
the opacity is initially very low and gradually increases as the
photon field builds-up, approaching its quasi-steady state value on
the dynamical time (by which time the emission episode leading to the
spike in the light curve is typically over). Furthermore, in this
model both the photon density and the typical value of $\theta_{12}$
decrease with radius along the trajectory of a test photon, further
reducing $\tau_{\gamma\gamma}$. 
%The time-averaging of the spectrum,
%and the dependence of the local value of $\tau_{\gamma\gamma}$ on the
%emission radius and angle lead to a non-trivial shape of the resulting
%spectral break \citep[e.g., Figure~8 and 13 of][]{Granot:08}.  This
%detailed analysis is based on a specific model, for which the value of
%the effective break energy is significantly higher than for the
%simpler and more approximate model discussed previously. 
For our discussion below we adopt the value of $\Gamma \sim 200$--$700$,
intermediate between the values inferred by the previous simple model
%and the more elaborate time-dependent model. 
and the time-dependent model for a very thin cooling layer.
The motivation for this
is that in some of the models discussed above the high-energy photons
are expected to be emitted from the bulk of the shocked region, rather
than from a thin cooling layer behind the shock front, in which case
such an intermediate value of the opacity might be expected.

The spectrum for time interval \emph{d} also has an extra
power-law component. It is much dimmer than that for time interval
\emph{c}, but its spectral index is similar.  The break 
feature is marginally significant. A straightforward interpretation of
this behavior is that the critical energy $E_c$ is larger than that
for interval \emph{c}. However, it is also possible that a
different emission component, which is responsible for the LAT
temporally extended emission at $T \gtrsim 20\;$s, contributes to the
high-energy emission in interval \emph{d}. This may hide a
possible spectral break of the extra component.

We emphasize that this burst is the first GRB that exhibits a spectral
break that can be used to estimate (as opposed to bound) its bulk
Lorentz factor, presuming that this feature is due to pair
production attenuation. Other LAT GRBs do not show any
clear high-energy spectral breaks \citep[a circumstance evinced in EGRET bursts with lower count 
statistics; see the overview in][]{baring06}.  
The lower limits of the bulk Lorentz factors for those GRBs have been derived by
$\tau_{\gamma\gamma}(E_h) < 1$, where $E_h$ is the highest photon
energy detected; $\Gamma \gtrsim 900, \Gamma \gtrsim 1200$, and
$\Gamma \gtrsim 1000$, for GRB~080916C, GRB~090510 and GRB~090902B,
respectively, using the simple model described above. In the
time-dependent thin-shell model of \citet{Granot:08} all of these
lower limits would be lower by about a factor of $\sim 3$.
%In addition, if the temporally extended GeV emissions of those bursts
%are from the external shock emission, then the early onsets of the
%afterglow emission are consistent with the initial bulk Lorentz
%factors of the ejecta as high as $\Gamma \gtrsim 1000$ (though in
%the stellar wind environment expected for the massive star progenitors
%of long GRBs a Lorentz factor of a few hundred is typically enough for
%the afterglow onset to occur near the end of the prompt GRB emission). 
Thus, the inferred $\Gamma \sim 200$--$700$ of GRB~090926A is smaller than the lower limits for 
other LAT GRBs.
%\citep[but typical of $\Gamma_{\rm min}$ inferred
%for EGRET bursts; e.g., see the overview in][] {baring06}.  
On the other hand, it is consistent with the constraints on $\Gamma$ for
other GRBs put by different methods:
%prior to the {\it Fermi} era: 
the observed broad flux peaks of some optical afterglows, typically $\sim
10^2$--$10^3\;$s after the burst triggers, appear to signify the
afterglow onset, and thus indicate $\Gamma \sim {\rm a~few} \times
100$ \citep{molinari07,oates09}; the possible thermal emission in some
bursts may suggest the photospheric radii of the jets, which indicate
$\Gamma \simeq 300$--$700$ \citep{peer07,ryde09}. 
These suggest that the Lorentz factors of GRBs are widely distributed over
a range $\gtrsim 10^2$.

\subsection{Temporally Extended Emission}

The emission in the LAT energy range lasts much longer than that in the GBM energy range,
and the flux in the LAT energy range shows a power-law decay, $F_{\nu} \propto (T-T_0)^{a}$ with
$a = -1.69 \pm 0.03$.
This behavior is similar to that seen in other LAT GRBs,
which have decay indices $a = -1.2 \pm 0.2$ for GRB~080916C \citep{2009Sci...323.1688A}, $a = -1.38 \pm 0.07$ for GRB~090510 \citep{depasquale09}, 
and $a \approx -1.5$ for GRB~090902B \citep{Fermi...GRB090902B}.
These may be explained as synchrotron emission from the external forward shock 
\citep{2009MNRAS.400L..75K,ghisellini09, depasquale09}.

For GRB~090926A, this interpretation seems consistent with the nearly constant spectral
index at $\gtrsim 20\;$s (Figure~6).  In this scenario, the starting time of the 
self-similar phase of the forward shock should be $\lesssim T_{\rm dur} \simeq 13\;$s.
This means that the ejecta is in the thick shell regime or in the borderline of the 
thick and thin shell regimes \citep{sari97,kobayashi03}, which corresponds to
$T_{\rm dec} = [(3-k)\mathcal{E}_{k,iso}/(2^{5-k}\pi A_{\rm ext} c^{5-k} \Gamma^{2(4-k)})]^{1/(3-k)}
(1+z) \lesssim T_{\rm dur} \simeq 13\;$s, where $\mathcal{E}_{k,iso}$ is the isotropic-equivalent kinetic
energy of the ejecta producing the external shock, and the external density profile is
defined as $n_{\rm ext} m_p = A_{\rm ext} R^{-k}$. 
This relation allows us to put a lower limit on the initial Lorentz factor of the ejecta.
For the uniform density case ($k=0$), $\Gamma \gtrsim 750\;
(\mathcal{E}_{\gamma,iso}/2\times10^{54}\;{\rm erg})^{1/8}
(\eta_{\gamma}/0.2)^{-1/8} (n_{\rm ext}/1\;{\rm cm}^{-3})^{-1/8}$, and for the wind medium case
($k=2$), $\Gamma \gtrsim 290\;(\mathcal{E}_{\gamma,iso}/2\times10^{54}\;{\rm erg})^{1/4}
(\eta_{\gamma}/0.2)^{-1/4} (A_{\rm ext}/5\times10^{11}\;{\rm g}\;{\rm cm}^{-1})^{-1/4}$,
where we define $\eta_{\gamma} = \mathcal{E}_{\gamma,iso}/\mathcal{E}_{k,iso}$,
and $A_{\rm ext} \simeq 5\times10^{11}\;{\rm g}\;{\rm cm}^{-1}$ corresponds to the 
value of typical Wolf-Rayet stars in our Galaxy, which have mass loss rates 
$\simeq 10^{-5} M_{\odot}\;{\rm yr}^{-1}$ and wind velocities $\simeq 1000\;{\rm km}\;{\rm s}^{-1}$.
\citet{cenko10} analyzed the observed late optical and 
X-ray afterglows of GRB~090926A and showed that they can be explained by synchrotron
emission from the adiabatic forward shock propagating into the wind medium with 
$A_{\rm ext} \simeq 3.4\times10^{11}\;{\rm g}\;{\rm cm}^{-1}$. This could provide 
$\Gamma \gtrsim 320 \;(\eta_{\gamma}/0.2)^{-1/4}$, 
which may be consistent with $\Gamma \sim 200$--$700$ of this burst
inferred by the pair attenuation opacity argument.
Note that the estimate $\Gamma \sim 200$--$700$ is only for the bulk Lorentz
factor of the shell emitting the spike in time interval \emph{c}, while the lower limit 
$\Gamma \gtrsim 320 (\eta_{\gamma}/0.2)^{-1/4}$ is relevant for the mean or typical value,
weighed over the energy in the whole outflow.

\section{Conclusions}
\label{sec:conclusions}
GRB~090926A is one of the brightest long bursts detected by the GBM and LAT instruments on \Fermi\ with high energy events up to $\sim 20\;$GeV. As in other bursts (GRB~090510, GRB~090902B), this burst shows an extra hard component in its integrated spectrum, but for the first time we significantly detect a spectral break around 1.4~GeV. The time-resolved spectral analysis shows that the extra component significantly dominates the emission in the high ($>1\;$MeV) energy range at the time of the narrow pulse which is simultaneously observed by LAT and GBM. At earlier times, the spectrum is described by a standard Band model while at later times the extra component is significant, but a spectral break feature is only marginally significant. Correlation between the lowest and highest energy light curves implies that the origins of the Band component and the extra power-law component are related around the time of the sharp pulse.

The $\sim 3.3\;$s delay of the LAT emission onset can be explained as the 
overall flux increase and the spectral hardening of the Band component,
since the clear emergence of the extra component occurs only at a later time.
However we may not exclude a contribution from the extra component in the early
times, whose flux is intrinsically just below the Band component or suppressed by 
a lower spectral cut-off.  The high temporal variability of the extra component and the correlation of the Band and extra components put strong constraints on the external shock scenario: the external medium needs to be highly clumpy, and the emission mechanisms of the two components should be related.

From the spectral break we have computed the bulk Lorentz factor of the emitting shell and find a range of $\Gamma \sim 200$--$700$, depending on the assumption of the homogeneity and time-dependence of the photon field, as well as on the assumption that the cutoff is due to the pair production attenuation. Even if we cannot distinguish between leptonic and hadronic emission for the extra component, we note that such a moderate Lorentz factor could alleviate the problem of the energy budget in hadronic emission models, as for GRB~090510. Comparison of this estimate with the large lower limits for other LAT GRBs and the estimates for GRBs that occurred before \Fermi\ may imply that the bulk Lorentz factors of GRBs are widely distributed over a range of values $\gtrsim 100$. In addition, the early deceleration of the forward shock inferred by the LAT temporal extended emission and the density of the external medium inferred by the late optical and X-ray afterglows \citep{cenko10} can put a lower limit on the bulk Lorentz factor of the entire shell just before the deceleration, $\Gamma \gtrsim 290 (\eta_\gamma/0.2)^{-1/4}$, which is consistent with the estimate of $\Gamma \sim 200$--$700$ for the region that corresponds to the emission around the spike.

Further LAT detections of bright GRBs will enable us to observe other bright extra components and constrain their origins and spectral breaks and their relation to excesses below 20~keV and the temporally extended emission.

\acknowledgments
We thank R. Mochkovitch for useful discussions.
The \Fermi\ GBM collaboration acknowledges support for GBM development,
operations and data analysis from NASA in the US and BMWi/DLR in Germany.
The \Fermi\ LAT Collaboration acknowledges support from a number of
agencies and institutes for both development and the operation of the
LAT as well as scientific data analysis. These include NASA and DOE in
the United States, CEA/Irfu and IN2P3/CNRS in France, ASI and INFN in
Italy, MEXT, KEK, and JAXA in Japan, and the K.~A.~Wallenberg
Foundation, the Swedish Research Council and the National Space Board
in Sweden. Additional support from INAF in Italy and CNES in France
for science analysis during the operations phase is also gratefully
acknowledged.

%\bibliographystyle{apj}
%\bibliography{GLAST_GRB}

%\if0

%\fi

%\clearpage
%\newcommand{\pfrac}[2]{\left(\frac{#1}{#2}\right)}
%\newcommand{\Epeak}{{E_{\rm peak}}}

\end{document}